\documentclass[10pt, two column, twoside]{IEEEtran}

\usepackage{amssymb}
\usepackage{amsmath}
\usepackage[lined,boxed,commentsnumbered, ruled]{algorithm2e}
\usepackage{mathrsfs}

\usepackage{pgfplots}
\usepackage{tikz}
\usetikzlibrary{arrows}
\usepackage{subfigure}
\usepackage{graphicx,booktabs,multirow}

\definecolor{colorhkust}{RGB}{20,43,140}
\definecolor{colortsinghua}{RGB}{116,52,129}
\definecolor{color1}{HTML}{D0B22B}

\newtheorem{theorem}{Theorem}

\newtheorem{proposition}{Proposition}

\newtheorem{remark}{Remark}

\newcommand{\bs}{\boldsymbol}
\newcommand{\tabincell}[2]{\begin{tabular}{@{}#1@{}}#2\end{tabular}}
\newcommand{\rev}{\color{black}}

\begin{document}

\title{Smoothed $L_p$-Minimization for Green Cloud-RAN with User Admission Control}
 \author{Yuanming~Shi,~\IEEEmembership{Member,~IEEE,}
 Jinkun Cheng,
 Jun~Zhang,~\IEEEmembership{Senior~Member,~IEEE,}\\
 Bo Bai,~\IEEEmembership{Member,~IEEE,} 
 Wei Chen,~\IEEEmembership{Senior~Member,~IEEE,}
 and~Khaled~B.~Letaief,~\IEEEmembership{Fellow,~IEEE}
 \thanks{Manuscript received xxx; revised xxx; accepted xxx. Date of publication xxx; date of current version xxx. This work is partially supported by the Hong Kong Research Grant Council under Grant No. 16200214, National Basic Research Program of China (973 Program) No. 2013CB336600, NSFC Excellent Young Investigator Award No. 61322111, Openning Research Funding of State Key Lab of Networking and Switching Technology, National Nature Science Foundation of China (NSFC) under Grant No. 61401249, and Specialized Research Fund for the Doctoral Program of Higher Education (SRFDP) under Grant No. 20130002120001.}
 \thanks{Y. Shi is with the School of Information Science and Technology, ShanghaiTech University, Shanghai, China (e-mail: shiym@shanghaitech.edu.cn).}
\thanks{ J. Zhang and K. B. Letaief are with the Department of Electronic and Computer Engineering,
 Hong Kong University of Science and Technology, Hong Kong (e-mail: \{eejzhang,
eekhaled\}@ust.hk).} 
\thanks{J. Cheng, B. Bai and W. Chen are with the Department of Electronic Engineering, Tsinghua University, Beijing, China (e-mail: cjk13@mails.tsinghual.edu.cn, \{eebobai, wchen\}@tsinghua.edu.cn).}}

\maketitle

\begin{abstract}
The cloud radio access network (Cloud-RAN) has recently been proposed as one cost-effective and energy-efficient technique for 5G wireless networks. By moving  the signal processing functionality to a single baseband unit (BBU) pool, centralized signal processing and resource allocation are enabled in Cloud-RAN, thereby providing the promise of improving the energy efficiency via effective network adaptation and  interference management. In this paper,  we propose a holistic sparse optimization framework to design green Cloud-RAN by taking into consideration the power consumption of the fronthaul links, multicast services, as well as user admission control. Specifically, we first identify the sparsity structures in the solutions of both the network power minimization and user admission control problems, which call for adaptive  
remote radio head (RRH) selection and user admission. However, finding the optimal sparsity structures turns out to be NP-hard, with the coupled challenges of the $\ell_0$-norm based objective functions and the nonconvex quadratic QoS constraints due to multicast beamforming. In contrast to the previous works on convex but non-smooth sparsity inducing approaches, e.g., the group sparse beamforming algorithm based on the mixed $\ell_1/\ell_2$-norm relaxation \cite{Yuanming_TWC2014}, we adopt the nonconvex but smoothed $\ell_p$-minimization ($0<p\le 1$) approach to promote sparsity in the multicast setting, thereby enabling efficient algorithm design based on the principle of the majorization-minimization (MM) algorithm and the semidefinite relaxation (SDR) technique. {\rev{In particular, an iterative reweighted-$\ell_2$ algorithm is developed, which will converge to a Karush-Kuhn-Tucker (KKT)
point of the relaxed smoothed $\ell_p$-minimization problem from the SDR technique}}. We illustrate the effectiveness of the proposed algorithms with extensive simulations for network power minimization and user admission control in multicast Cloud-RAN.         
\end{abstract}


\section{Introduction}
The great success of wireless industry is  driving the proposal of new services and innovative applications, such as Internet of Things (IoT) and mobile Cyber-Physical applications, which yield an exponential growth of wireless traffic with billions of connected devices. To handle the enormous mobile data traffic, network densification and heterogeneity supported by various radio access technologies (e.g., massive MIMO \cite{Rusek_SPM2013} and millimeter-wave communications \cite{Rappaport_IEEEPrecoding2014}) have become an irreversible trend in 5G wireless networks \cite{Jeff_JSAC5G}. However, this will have a profound impact and bring formidable challenges to the design of 5G wireless communication systems  in terms of energy efficiency, capital expenditure (CAPEX), operating expenditure (OPEX), and interference management \cite{Yuanming_WCMLargeCVX}. In particular, the energy consumption will become prohibitively high in such dense wireless networks in the era of mobile data deluge. Therefore,  to accommodate the upcoming diversified and high-volume data services in a cost-effective and energy-efficient way, a paradigm shift is required in the design of 5G wireless networks.

By leveraging the cloud computing technology \cite{Wubben_SPM2014Cloud}, the cloud radio access network (Cloud-RAN) \cite{mobile2011c,Peng_HCRAN} is a disruptive technology to address the key challenges of energy efficiency in 5G wireless networks. Specifically, by moving the baseband units (BBUs) into a single BBU pool (i.e., a cloud data center) with shared computation resources, scalable and parallel signal processing, coordinated resource allocation and cooperative interference management algorithms \cite{Jun_2009networkedTWC,Gesbert_JSAC10} can be enabled among a large number of radio access points, thereby significantly improving the energy efficiency  \cite{Yuanming_TWC2014, Rui_TWC2015GSBF} and spectral efficiency \cite{Yuanming_LargeSOCP2014}. As the conventional compact base stations are replaced by low-cost and low-power remote radio heads (RRHs), which are connected to the BBU pool through high-capacity and low-latency fronthaul links, Cloud-RAN provides a cost-effective and energy-efficient way to densify the radio access networks \cite{Yuanming_WCMLargeCVX}. 

While Cloud-RAN has a great potential to reduce the energy consumption of each RRH, with additional fronthaul link components and dense deployment of RRHs, new challenges arise for designing green Cloud-RAN. In particular, instead of only minimizing the total transmit power consumption via coordinated beamforming \cite{WeiYu_WC10}, the network power consumption consisting of the fronthaul link power consumption and the RRH power consumption should be adopted as the performance metric for designing green Cloud-RAN  \cite{Yuanming_TWC2014,Rui_TWC2015GSBF,Yuanming_SDP2014}. To minimize the network power consumption, a group sparse beamforming framework was proposed in \cite{Yuanming_TWC2014} to  adaptively select the active RRHs and the corresponding fronthaul links via controlling the group sparsity structures of the beamforming vectors. Such an idea of exploiting sparsity structures in the solutions has also demonstrated its effectiveness in solving other mixed combinatorial optimization problems in Cloud-RAN, e.g., the data assignment
problem  \cite{Wei_IA2014Sparse} and the joint uplink and downlink network power minimization problem \cite{Rui_TWC2015GSBF}.

Although network adaption by selecting the active RRHs provides a promising way to minimize the network power consumption, it is critical to maximize the user capacity (i.e., the number of admitted users) when the network power minimization problem is infeasible \cite{Luo_TSP2009efficient,Tan_JSAC2014,Tony_2015heterogeneous}. This infeasibility
issue may often occur in the scenarios with a large number of mobile devices requesting high data rates  or some users with unfavorable channel conditions. Furthermore, exploiting the benefits of integrating diversified  unicast and multicast
services \cite{Schaefer_SPM2014Physcial} has been well recognized as  a promising way to improve the energy efficiency and user capacity, and thus multicast beamforming should be incorporated in Cloud-RAN. From the system design perspective,  to design a green Cloud-RAN with multicast transmission, a holistic approach  is needed to enable network adaptation for RRH selection and user admission in a unified way.

Unfortunately, such design problems fall into the category of highly complicated mixed combinatorial optimization problems. The key observation to {\rev{address}} this challenge is that the network power minimization and user admission control can be achieved by adaptively selecting the active RRHs and admitting the mobile users (MUs) via controlling the sparsity structures in the corresponding solutions. Specifically, for the network power minimization problem, selecting active RRHs is equivalent to controlling the group sparsity structure in the aggregative multicast beamforming vector \cite{Yuanming_TWC2014}. That is, all the beamforming coefficients of a particular RRH that is switched off need to  be set to zeros simultaneously. For the user admission control problem that is needed when the network power minimization problem is infeasible, maximizing the number of admitted users is equivalent to minimizing the number of violated QoS constraints \cite{boyd2004convex}. Mathematically, this is the same as minimizing the sparsity of the auxiliary vector indicating the violations of the QoS constraints.  Based on these observations, we will thus formulate both design problems as sparse optimization problems based on the $\ell_0$-norm minimization in a unified framework, based on which efficient algorithms will be developed.

\subsection{Contributions}  
Based on the above discussions, we propose a sparse optimization framework to design a multicast green Cloud-RAN as shown in Fig. {\ref{framework}}, thereby enabling adaptive RRH selection and user admission. However, in contrast to the previous works on the multicast beamforming problem \cite{Luo_2006transmitmulticasting} with convex objectives but nonconvex QoS constraints and the group sparse beamforming problem \cite{Yuanming_TWC2014} with convex QoS constraints but nonconvex objective functions in unicast Cloud-RAN, to design efficient algorithms for the resulting sparse optimization problems in multicast Cloud-RAN, we need to address the following coupled challenges in a unified way:
\begin{itemize}
\item Nonconvex quadratic QoS constraints due to multicast transmission;
\item Combinatorial objective functions for RRH selection and user admission.  
\end{itemize}  
In particular, we summarize the major contributions as follows:
\begin{enumerate}
\item A sparse optimization framework based on the $\ell_0$-norm minimization is proposed to design a multicast green Cloud-RAN by adaptive RRH selection and user admission via controlling the sparsity structures of the solutions. 

\item To address the combinatorial challenges in the objective functions, we propose a nonconvex but smoothed $\ell_p$-minimization approach to induce the sparsity structures in the solutions. {\rev{The main advantage of this method is that it helps develop the
group sparse inducing penalty with quadratic forms in the multicast beamforming vectors.
Therefore, the objective function in the resulting group sparse inducing
optimization problem is compatible with the nonconvex quadratic QoS constraints.
The SDR technique can then be adopted to solve the resulting nonconvex
quadratic group sparse inducing optimization problem.}} 
\item To address the challenges of the  nonconvex  smoothed objective functions and the nonconvex quadratic QoS constraints, we propose an iterative reweighted-$\ell_2$ algorithm to solve the resulting nonconvex smoothed $\ell_p$-minimization problems based on the principle of the MM algorithm and the SDR technique. {\rev{This algorithm is proven to converge to a KKT point of the relaxed smoothed $\ell_p$-minimization problems over the convex constraint set using the SDR technique}}.      

\item Simulation results will demonstrate the effectiveness of the proposed algorithms to minimize the network power minimization and maximize the user capacity, and their performance gains compared with the existing convex approximation approaches. In particular, the proposed algorithms can achieve near-optimal performance in the simulated settings.       
\end{enumerate}

\subsection{Related Works}
Sparse optimization by exploiting sparsity structures of the
solutions  has been proven to be very powerful to solve
various hard optimization problems in machine learning, compressive sensing and
high-dimensional statistics \cite{Bach_ML2011}. This approach has recently received enormous attentions in designing wireless networks, i.e., the group sparse beamforming approach for network adaption \cite{Yuanming_TWC2014,Rui_TWC2015GSBF,Z.Q.Luo_JSAC2013}, and data assignment in wireless backhaul networks \cite{Wei_IA2014Sparse, TonyQ.S._WC2013}. In particular, the convex relaxation approaches, e.g., the $\ell_1$-minimization \cite{Tan_JSAC2014}, the mixed $\ell_1/\ell_2$-norm
\cite{Yuanming_TWC2014} and the mixed $\ell_1/\ell_{\infty}$-norm \cite{Mehanna_SP2013}, have become popular due to the computational efficiency as well as  performance guarantees in some scenarios \cite{Tao_IT06}. 

To further improve the performance, there has been a great interest in applying nonconvex approaches in sparse optimization \cite{Ye_MP2011lp,Chartrand2008restricted,Daubechies_2010iteratively,Boyd_2008enhancing} by enhancing sparsity. In particular, it is observed that the nonconvex $\ell_p$-minimization approach performs better than the traditional convex $\ell_1$-minimization especially when the underlying model is very sparse \cite{Chartrand2008restricted}. Motivated by this result, we adopt the $\ell_p$-minimization approach to closely approximate the resulting $\ell_0$-norm based sparse optimization problems for multicast green Cloud-RAN. Furthermore, to  deal with the unique challenges with the
coupled nonconvex constraints and combinatorial objectives, thereby enabling efficient algorithm design, we will use a smoothed version of the $\ell_p$-minimization approach to induce the sparsity structures in the solutions. Note that the existing work on the group sparse beamforming \cite{Yuanming_TWC2014} can only handle problems with a combinatorial objective and convex seconder-order cone QoS constraints in unicast Cloud-RAN, and thus cannot be directly applied in the setting of multicast Cloud-RAN with nonconvex QoS constraints.

\subsection{Organization}
The remainder of the paper is organized as follows.  Section {\ref{sysmod}} presents the system model and problem formulations. Section {\ref{lpf}} presents an algorithmic framework for network power minimization and user admission control based on the smoothed $\ell_p$-minimization. The iterative reweighted-$\ell_2$ algorithm is developed in Section {\ref{em}}. Simulation results will be demonstrated in Section {\ref{simres}}. Finally, conclusions and discussions are presented in Section {\ref{confu}}.

\section{System Model and Problem Formulation}
\label{sysmod}
In this section, we first introduce the system model of the multicast Cloud-RAN. Then, the network power minimization problem is formulated. For the scenario when it is not feasible to serve all the MUs, the user admission control problem is formulated. It will be revealed that both problems are sparse optimization problems, for which unique challenges will be identified.

\subsection{System Model}
\begin{figure}[t]
  \centering
  \includegraphics[width=0.95\columnwidth]{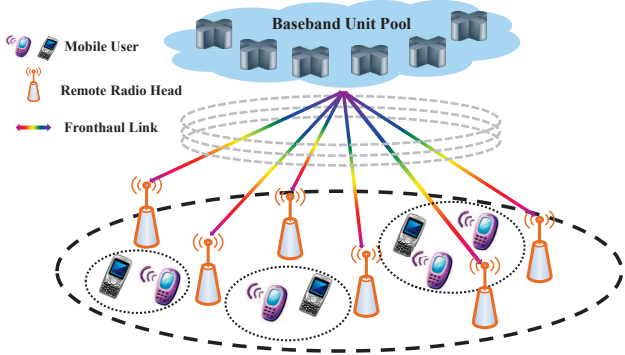}
 \caption{The architecture of multicast Cloud-RAN, in which, all the RRHs are connected to a BBU pool through high-capacity and low-latency optical fronthaul links. All the MUs in the same dashed circle form a multicast group and request the same message. To enable full cooperation among all the RRHs, it is assumed that all the user data and channel state information (CSI) are available at the BBU pool.}
 \label{cloudran}
 \end{figure}
We consider a multicast Cloud-RAN with $L$ multi-antenna RRHs and $K$ single-antenna MUs, where the $l$-th RRH is equipped with $N_l$ antennas, as shown in Fig. {\ref{cloudran}}. Let $\mathcal{L}=\{1,\dots, L\}$ and $\mathcal{N}=\{1,\dots, K\}$ denote
the sets of all the RRHs and all the MUs, respectively. We assume that the $K$ MUs form $M$ non-overlapping and non-empty multicast groups, which are denoted as $\{\mathcal{G}_1, \mathcal{G}_2,\dots, \mathcal{G}_M\}$ with $\cup_{i}\mathcal{G}_i=\mathcal{N}$
and $\mathcal{G}_j\cap\mathcal{G}_j=\emptyset$ with $\mathcal{G}_m$ as the set of MUs in the $m$-th multicast group.  Let $\mathcal{M}=\{1,\dots, M\}$ be the set of all the multicast groups. We consider the downlink transmission, and the centralized signal processing is performed at the BBU pool \cite{Yuanming_TWC2014, mobile2011c}.  

The propagation channel from the $l$-th RRH to the $k$-th MU is denoted as ${\bf{h}}_{kl}\in\mathbb{C}^{N_{l}}, \forall k, l$. Let ${\bf{v}}_{lm}\in\mathbb{C}^{N_{l}}$ be the transmit beamforming vector from the $l$-th RRH to MUs in multicast group $\mathcal{G}_m$. Then the transmit signal at the $l$-th RRH is given by
\setlength\arraycolsep{1.5pt}
\begin{eqnarray}
{\bf{x}}_l=\sum_{m=1}^{M}{\bf{v}}_{lm}s_m, \forall l\in\mathcal{L},
\end{eqnarray}
where $s_{m}\in\mathbb{C}$ is the encoded information symbol for the multicast
group $m$ with $\mathbb{E}[|s_{m}|^2]=1$. Assume that all the  RRHs have their own transmit power constraints, which form the following feasible set of the beamforming vectors ${\bf{v}}_{lm}$'s,
\begin{eqnarray}
\label{tp}
\mathcal{V}=\left\{{\bf{v}}_{lm}\in\mathbb{C}^{N_l}:\sum_{m=1}^{M}\|{\bf{v}}_{lm}\|_2^2 \le P_l, \forall l\in\mathcal{L}\right\},
\end{eqnarray}
where $P_l>0$ is the maximum transmit power of the $l$-th RRH.

 The received signal $y_{km}\in\mathbb{C}$ at MU $k$ in the $m$-th multicast group is given by

\begin{eqnarray}
y_{km}=\sum_{l=1}^{L}{\bf{h}}_{kl}^{\sf{H}}{\bf{v}}_{lm}s_{m}+\!\!\sum_{i\ne m}\sum_{l=1}^{L}{\bf{h}}_{kl}^{\sf{H}}{\bf{v}}_{li}s_{i}+n_{k}, \forall k\in\mathcal{G}_m,
\end{eqnarray}
where $n_{k}\sim\mathcal{CN}(0, \sigma_{k}^2)$
is the additive Gaussian noise at MU $k$. We assume that $s_{m}$'s and $n_{k}$'s
are mutually independent and all the users apply single user detection. Therefore, the signal-to-interference-plus-noise ratio (SINR) of MU $k$ in multicast group $m$ is given by
\begin{eqnarray}
\label{sinr}
{{\Gamma}}_{k,m}({\bf{v}})={{{\bf{v}}_m^{\sf{H}}{\bf{\Theta}}_k{\bf{v}}_m}\over{\sum_{i\ne m}{\bf{v}}_i^{\sf{H}}{\bf{\Theta}}_k{\bf{v}}_i}+\sigma_k^2}, \forall k\in\mathcal{G}_m,
\end{eqnarray}
where ${\bf{\Theta}}_k={\bf{h}}_k{\bf{h}}_k^{\sf{H}}\in\mathbb{C}^{N\times N}$ with $N=\sum_{l=1}^{L}N_l$ and ${\bf{h}}_k=[{\bf{h}}_{k1}^{\sf{H}}, {\bf{h}}_{k2}^{\sf{H}}, \dots, {\bf{h}}_{kL}^{\sf{H}}]^{\sf{H}}\in\mathbb{C}^{N}$
as the channel vector consisting of the channel
coefficients from all the RRHs to MU $k$, ${\bf{v}}_m=[{\bf{v}}_{1m}^{\sf{H}},
{\bf{v}}_{2m}^{\sf{H}},\dots,
{\bf{v}}_{Lm}^{\sf{H}}]^{\sf{H}}\in\mathbb{C}^{N}$ is the beamforming vector
consisting of the beamforming coefficients from all the RRHs
to the $m$-th multicast group, and ${\bf{v}}=[\tilde{\bf{v}}_l]_{l=1}^{L}\in\mathbb{C}^{MN}$ is the aggregative beamforming vector with $\tilde{\bf{v}}_l=[{\bf{v}}_{lm}]_{m=1}^{M}\in\mathbb{C}^{MN_l}$ as the beamforming vector consisting of all the beamforming coefficients from the $l$-th RRH to all the multicast groups.

\subsection{Network Power Minimization for Green Cloud-RAN}
With densely deployed RRHs, it is critical to enable energy-efficient transmission via centralized signal processing at the BBU pool. Coordinated beamforming among RRHs will help reduce the transmit power. Due to the mobile data traffic variations in temporal and spatial domains, it is also critical
to enable network adaptation to switch off some RRHs and the corresponding fronthaul links to save power \cite{Yuanming_TWC2014,Yang_TWC2014BS}. Thus, we need to consider the network power consumption when designing green Cloud-RAN, which is defined as the following  {combinatorial composite function} parameterized by the beamforming coefficients \cite{Yuanming_TWC2014},
\begin{eqnarray} 
p({\bf{v}})=p_1({\bf{v}})+p_2({\bf{v}}),
\end{eqnarray} 
with the total transmit power consumption, denoted as $p_1({\bf{v}})$, and the total relative fronthaul links power consumption, denoted as $p_{2}({\bf{v}})$,
given as
\begin{eqnarray}
\label{tp1}
p_1({\bf{v}})=\sum_{l=1}^{L}\sum_{m=1}^{M}{1\over{\eta_{l}}}\|{\bf{v}}_{lm}\|_2^2,
\end{eqnarray}
and 
\begin{eqnarray}
\label{fh1}
p_2({\bf{v}})=\sum_{l=1}^{L}P_l^{c} I({\rm{Supp}}({\bf{v}})\cap\mathcal{V}_l\ne\emptyset),
\end{eqnarray}
respectively. Here, $I({\rm{Supp}}({\bf{v}})\cap\mathcal{V}_l\ne\emptyset)$ is an indicator function that takes value zero if ${\rm{Supp}}({\bf{v}})\cap\mathcal{V}_l=\emptyset$ (i.e., all the beamforming coefficients at the $l$-th RRH are zeros, indicating that the corresponding RRH is switched off) and one otherwise, where $\mathcal{V}_l$ is defined as $\mathcal{V}_l:=\{M\sum_{l=1}^{L-1}N_l+1,\dots, M\sum_{l=1}^{L}N_l\}$, and ${\rm{Supp}}({\bf{v}})$ is the support of the vector $\bf{v}$. In (\ref{tp1}), $\eta_{l}>0$ is the drain inefficiency coefficient
of the radio frequency power amplifier with the typical value as $25\%$, and $P_{l}^c\ge 0$ in (\ref{fh1}) is the relative fronthaul link power consumption \cite{Yuanming_TWC2014}, which is the static power saving when both the RRH and the corresponding fronthaul link are switched off. {\rev{For the passive optical fronthaul network \cite{Dhaini_ITN2013}, $P_l^c$ is given by $(P_{a,l}^{\textrm{rrh}}+P_{a,l}^{\textrm{fn}})-(P_{s,l}^{\textrm{rrh}}+P_{s,l}^{\textrm{fn}})$ with $P_{a,l}^{\textrm{rrh}}$ $(P_{s,l}^{\textrm{rrh}})$ and $P_{a,l}^{\textrm{fn}}$ $(P_{s,l}^{\textrm{fn}})$ as the power consumptions for the $l$-th RRH and the $l$-th fronthaul link in the active (sleep) mode, respectively. The typical values are $P_{a,l}^{\textrm{rrh}}=6.8 W$, $P_{s,l}^{\textrm{rrh}}=4.3W$, $P_{a,l}^{\textrm{fn}}=3.85W$, $P_{s,l}^{\textrm{fn}}=0.75W$ and $P_l^c=5.6W$ \cite{Yuanming_TWC2014, Dhaini_ITN2013}. \emph{Note that the energy consumption of the optical fronthaul links should depend on the receiving periods and data transmission \cite{Dhaini_ITN2013}, which is a function
of the beamforming vectors}.}} Given the target SINR requirements $(\gamma_1,\gamma_2,\dots, \gamma_K)$ for all the MUs, to design a multicast green Cloud-RAN, we propose to minimize the network power consumption subject to the QoS constraints and the RRH transmit power constraints. Specifically, we have the following QoS constraints
\begin{eqnarray}
{\Gamma}_{k,m}({\bf{v}})\ge \gamma_k, \forall k\in\mathcal{G}_m, m\in\mathcal{M},
\end{eqnarray}
which can be rewritten as the following quadratic constraints
\begin{eqnarray}
\label{si}
\!\!\! F_{k,m}({\bf{v}})=\gamma_k\left(\sum\nolimits_{i\ne m}{\bf{v}}_i^{\sf{H}}{\bf{\Theta}}_k{\bf{v}}_i+\sigma_{k}^2\right)-{\bf{v}}_m^{\sf{H}}{\bf{\Theta}}_k{\bf{v}}_m\le
0,
\end{eqnarray}
for any $k\in\mathcal{G}_m$ and $m\in\mathcal{M}$, which are nonconvex. Therefore,  the network power minimization problem can be formulated as\begin{eqnarray}
\label{pm}
\mathop {\textrm{minimize}}_{{\bf{v}}\in \mathcal{V}}&&p_{1}({\bf{v}})+p_2({\bf{v}})\nonumber\\
\textrm{subject to}&& F_{k,m}({\bf{v}})\le 0,
\forall
k\in\mathcal{G}_m, m\in\mathcal{M},
\end{eqnarray}
which is highly intractable due to the nonconvex combinatorial composite objective and the nonconvex quadratic QoS constraints (\ref{si}). When there are some RRHs needed to be switched off to minimize the network power consumption, the  solution of problem ({\ref{pm}}) has the \emph{group sparsity} structure \cite{Yuanming_TWC2014}. That is, all the beamforming coefficients in $\tilde{\bf{v}}_l$, which forms a group at the $l$-th RRH, are set to be zeros simultaneously if the $l$-th RRH needs to be switched off. 

Therefore, inspired by the fact that the solution of problem (\ref{pm})
has the group
sparsity structure in the aggregative beamforming vector $\bf{v}$, the weighted mixed $\ell_1/\ell_2$-norm was proposed in \cite{Yuanming_TWC2014} to relax the combinatorial composite function as the tightest convex surrogate to induce the group sparsity structure in the solution $\bf{v}$ to guide the RRH selection, defined as
\begin{eqnarray}
\label{mn}
\mathcal{J}({\bf{v}})=\sum_{l=1}^{L}\rho_{l}\|\tilde{\bf{v}}_l\|_2,
\end{eqnarray}
where $\rho_l>0$ is the weight for the beamforming coefficients
group $\tilde{\bf{v}}_l$ at RRH $l$.  

To handle the nonconvex QoS constraints in (\ref{si}), we propose to lift the problem to higher dimensions with variables ${\bf{Q}}_m={\bf{v}}_m{\bf{v}}_m^{\sf{H}}\in\mathbb{C}^{N\times N}, \forall m$, which will help to apply the semidefinite relaxation (SDR) technique. However, we cannot extract the variables  ${\bf{Q}}_m$'s from the non-smooth mixed $\ell_1/\ell_2$-norm (\ref{mn}). Therefore,  this convex relaxation approach cannot
be directly applied to solve the network power minimization problem in  multicast Cloud-RAN. Instead, to leverage the SDR technique, we need to develop a new group sparsity inducing approach with quadratic forms of the beamforming vectors, which will be presented in Section {\ref{lpf}} and form one major contribution of this paper.    

\subsection{User Admission Control}
With QoS constraints for potentially a large number of MUs in the serving area, it may happen that the network power minimization problem (\ref{pm}) is infeasible from time to time. In such scenarios, the design problem will be  to maximize the user capacity (i.e., the maximum number of MUs that can be supported) via user admission control \cite{Luo_TSP2009efficient,Tan_JSAC2014}, while serving these MUs with the minimum transmit power.
While this aspect is ignored in \cite{Yuanming_TWC2014}, it is critical for practical applications. Mathematically, by adding  auxiliary variables $x_k$'s to the right-hand side of the corresponding inequalities in (\ref{si}), to maximize the number of admitted MUs is  equivalent
to minimize the number of non-zero $x_k$'s \cite[Section 11.4]{boyd2004convex}.
Therefore, the user admission control problem can be formulated as the following sparsity minimization problem, 
\begin{eqnarray}
\label{ua}
\mathop {\textrm{minimize}}_{{\bf{v}}\in \mathcal{V}, {\bf{x}}\in\mathbb{R}_{+}^K}&&\|{\bf{x}}\|_{0}\nonumber\\
\textrm{subject to}&&F_{k,m}({\bf{v}})\le x_k, \forall k\in\mathcal{G}_m,
m\in\mathcal{M},
\end{eqnarray}
where $\mathbb{R}_{+}^{K}$ represents the $K$-dimensional nonnegative real
vector. Once the admitted MUs are determined, coordinated beamforming can then be applied to minimize the total transmit power. The solution ${\bf{x}}=[x_k]$ of problem (\ref{ua}) has the {\emph{individual sparsity}} structure, i.e., $x_k=0$ indicates that the $k$-th MU can be admitted. Therefore, the sparsity level will be increased if more MUs can be admitted. 

Observing that both the sparse optimization problems (\ref{pm}) and (\ref{ua}) possess the same structure with nonconvex quadratic constraints and combinatorial objectives, in this paper, we will propose a unified way to {\rev{handle}} them based on a smoothed $\ell_p$-minimization approach.

\subsection{Problem Analysis}
In this subsection, we analyze the unique challenges of the  network power minimization problem
(\ref{pm}) and the user admission control problem (\ref{ua})  in the context of multicast Cloud-RAN. In particular, the differences from the previous works \cite{Yuanming_TWC2014,Rui_TWC2015GSBF} will be highlighted.
\subsubsection{Nonconvex Quadratic Constraints} 
\label{sdr}
The physical-layer multicast beamforming problem \cite{Luo_2006transmitmulticasting} yields nonconvex quadratic QoS constraints (\ref{si}), while only unicast services were considered in \cite{Yuanming_TWC2014,Rui_TWC2015GSBF}. The SDR technique \cite{Luo_2008quality} proves to be an effective way to obtain good approximate solutions to these problems by lifting the quadratic constraints into higher dimensions, which will also be adopted in this paper.  

\subsubsection{Combinatorial  Objective Functions}

Although the SDR technique provides an efficient way to convexify the nonconvex quadratic QoS constraints in problems (\ref{pm}) and (\ref{ua}), the inherent combinatorial  objective functions still make the resulting problems highly intractable. While the  $\ell_1$-norm can be adopted to relax the
$\ell_0$-norm in  problem (\ref{ua}) after SDR, which is also known as the sum-of-infeasibilities relaxation in
optimization theory \cite{Tan_JSAC2014,boyd2004convex}, the convex relaxation approach based on the non-smooth mixed $\ell_1/\ell_2$-norm \cite{Yuanming_TWC2014} cannot be applied to problem (\ref{pm}), as we cannot extract the variables ${\bf{Q}}_m$'s after SDR.

Therefore, in this paper, we propose a new powerful approach to induce the sparsity structures in the solutions for both problems (\ref{pm}) and (\ref{ua}), which is based on a smoothed $\ell_p$-norm \cite{Ye_MP2011lp}. The main advantage of this method is that it can help develop group sparsity inducing penalties with quadratic forms in the beamforming vectors, thereby leveraging the SDR technique to relax the nonconvex quadratic QoS constraints for problem (\ref{pm}). Furthermore, by adjusting the parameter $p$, this approach has the potential to yield a better approximation for the original $\ell_0$-norm based objectives, thereby providing improved solutions for problems (\ref{pm}) and (\ref{ua}). The smoothed $\ell_p$-minimization framework will be presented in Section {\ref{lpf}}, while the iterative reweighted-$\ell_2$  algorithm will be developed in Section {\ref{em}} to solve the smoothed $\ell_p$-minimization problem.

\section{A Smoothed $\ell_p$-Minimization Framework for Network Power Minimization with User Admission Control}
\label{lpf}

In this section, we first present the  smoothed $\ell_p$-minimization method as a unified way to  induce sparsity structures in the solutions of problems (\ref{pm}) and (\ref{ua}), thereby providing guidelines for RRH selection and user admission. After obtaining the active RRHs and admitted MUs by performing the corresponding selection procedure, we will minimize the total transmit power consumption for the size-reduced network. The algorithmic advantages and performance improvement of the proposed smoothed $\ell_p$-minimization based framework will be revealed in the sequel.

\subsection{Smoothed $\ell_p$-Minimization for Sparsity Inducing} 
To promote sparse solutions, instead of applying the convex $\ell_1$-minimization approach, we adopt a nonconvex $\ell_p$-minimization $(0<p\le 1)$ approach   to seek a tighter approximation of the   $\ell_0$-norm in the objective functions in problems (\ref{pm}) and (\ref{ua}) \cite{Ye_MP2011lp}. {\rev{This is motivated by the fact that the $\ell_0$-norm $\|{\bs{z}}\|_0$ is the limit as $p\rightarrow 0$ of $\|{\bs{z}}\|_p^p$
in the sense of $\|{\bs{z}}\|_0=\lim_{p\rightarrow 0} \|{\bs{z}}\|_p^p=\lim_{p\rightarrow
0}\sum|{z}_i|^p$. We thus adopt $\|{\bf{x}}\|_p^p$ as the optimization objective function to seek sparser solutions}}. Furthermore, to enable efficient algorithm design as well as induce the quadratic forms in the resulting approximation problems, we instead adopt the following smoothed version of $\|{\bs{z}}\|_p^p$ to induce sparsity:
\begin{eqnarray}
\label{smp}
f_{p}({\bs{z}}; \epsilon):=\sum_{i=1}^{m}(z_i^2+\epsilon^2)^{p/2},
\end{eqnarray}
for  ${\bs{z}}\in\mathbb{R}^{m}$ and some small fixed regularizing parameter $\epsilon>0$. Based on the smoothed $\ell_p$-norm (\ref{smp}), we will present the algorithmic advantages of the smoothed $\ell_p$-minimization approach in Section {\ref{em}}.

\subsubsection{Smoothed $\ell_p$-Minimization for Group Sparsity Inducing}
For network power minimization, to seek quadratic
forms of beamforming vectors in
the objective functions to leverage the SDR technique for the  non-convex quadratic QoS constraints, we adopt the smoothed $\ell_p$-norm $f_p({\bs{z}}; \epsilon)$ (\ref{smp}) to induce group sparsity in the aggregative beamforming vector $\bf{v}$ for problem (\ref{pm}), resulting the following optimization problem:\begin{eqnarray}
\label{qcqp}
\mathop {\textrm{minimize}}_{{\bf{v}}\in \mathcal{V}}&&\sum_{l=1}^{L}\rho_l
(\|\tilde{\bf{v}}_l\|_2^2+\epsilon^2)^{p/2}\nonumber\\
\textrm{subject to}&& F_{k,m}({\bf{v}})\le 0,
\forall
k\in\mathcal{G}_m, m\in\mathcal{M}.
\end{eqnarray}
The induced (approximated) group sparse beamformers will guide the RRH selection. The resulting problem (\ref{qcqp}) thus becomes a quadratic optimization problem and enjoys the algorithmic advantages.  

\subsubsection{Smoothed $\ell_p$-Minimization for User Admission Control}
For user admission control, we adopt the smoothed  $\ell_p$-norm (\ref{smp}) to
approximate the objective function in problem (\ref{ua}), yielding the following optimization problem:
\begin{eqnarray}
\label{qcqp2}
\mathop {\textrm{minimize}}_{{\bf{v}}\in \mathcal{V}, {\bf{x}}\in\mathbb{R}_{+}^K}&&\sum_{k=1}^{K}(x_k^2+\epsilon^2)^{p/2}\nonumber\\
\textrm{subject to}&&F_{k,m}({\bf{v}})\le x_k, \forall k\in\mathcal{G}_m,
m\in\mathcal{M}.
\end{eqnarray}
This will help to induce individual sparsity in the auxiliary variables ${\bf{x}}$, thereby guiding the user admission. 

Although the resulting optimization problems (\ref{qcqp}) and (\ref{qcqp2}) are still nonconvex, they can readily be solved by the SDR technique and the MM algorithm. Specifically, we will demonstrate that the nonconvex quadratic QoS constraints can be convexified by the SDR technique in the next subsection. The resulting convex constrained smoothed $\ell_p$-minimization problem will be solved by the MM algorithm, yielding an iterative reweighted-$\ell_2$ algorithm, as will be presented in Section {\ref{em}}.

\subsection{SDR for Nonconvex Quadratic Constraints}
In this part, we will demonstrate how to apply the SDR technique to resolve the challenge of the nonconvex quadratic QoS constraints in both problems (\ref{qcqp}) and (\ref{qcqp2}). Specifically, let ${\bf{Q}}_m={\bf{v}}_m^{\sf{H}}{\bf{v}}_m\in\mathbb{C}^{N\times
N}$ with ${\rm{rank}}({\bf{Q}}_m)=1, \forall m\in\mathcal{M}$. Therefore,
the QoS constraint (\ref{si}) can be rewritten as
\begin{eqnarray}
L_{k,m}({\bf{Q}})\le 0, \forall k\in\mathcal{G}_m, m\in\mathcal{M},
\end{eqnarray}
with $L_{k,m}({\bf{Q}})$ given by  
\begin{eqnarray}
\!\!\!\!L_{k,m}({\bf{Q}})=\gamma_k \left(\sum\limits_{i\ne
m}{\rm{Tr}}({\bf{\Theta}}_k{\bf{Q}}_i)+\sigma_{k}^2\right)-{\rm{Tr}}({\bf{\Theta}}_k{\bf{Q}}_m),
\end{eqnarray}
where ${\bf{Q}}=[{\bf{Q}}_m]_{m=1}^{M}$ and all  ${\bf{Q}}_m$'s are rank-one
constrained. The per-RRH transmit power constraints (\ref{tp}) can  be rewritten as 
\begin{eqnarray} 
\mathcal{Q}=\left\{{\bf{Q}}_{m}\in\mathbb{C}^{N}:\sum_{m=1}^{M}{\rm{Tr}}({\bf{C}}_{lm}{\bf{Q}}_m)\le
P_l, \forall l\in\mathcal{L}\right\},
\end{eqnarray}
where ${\bf{C}}_{lm}\in\mathbb{R}^{n\times
n}$ is a block diagonal
matrix with the identity matrix ${\bf{I}}_{N_l}$ as the $l$-th main diagonal
block square matrix and zeros elsewhere.

Therefore, by dropping the rank-one constraints for all the matrices ${\bf{Q}}_m$'s
based on the
principle of the SDR technique, problem (\ref{qcqp}) can be
relaxed as
\begin{eqnarray}
\label{gsid}
\mathscr{P}: \mathop {\textrm{minimize}}_{{\bf{Q}}\in\mathcal{Q}}&&
\sum_{l=1}^{L}\rho_l\left(\sum_{m=1}^{M}{\rm{Tr}}({\bf{C}}_{lm}{\bf{Q}}_m)+\epsilon^2\right)^{p/2}
 \nonumber\\
\textrm{subject to}&&L_{k,m}({\bf{Q}})\le 0, \forall k\in\mathcal{G}_m,
\nonumber\\
&& {\bf{Q}}_m\succeq {\bf{0}}, \forall m\in\mathcal{M}.
\end{eqnarray}

Similarly, by dropping the rank-one constraints of all the
matrices ${\bf{Q}}_m$'s, problem (\ref{qcqp2}) can be relaxed as
\begin{eqnarray}
\label{indsp}
\mathscr{D}:\mathop {\textrm{minimize}}_{{\bf{Q}}\in\mathcal{Q},
{\bf{x}}\in\mathbb{R}_{+}^K}&& \sum_{k=1}^{K}(x_k^2+\epsilon^2)^{p/2}\nonumber\\
\textrm{subject to}&&L_{k,m}({\bf{Q}})\le x_k,
\forall k\in\mathcal{G}_m,
\nonumber\\
&&{\bf{Q}}_m\succeq {\bf{0}}, \forall m\in\mathcal{M}.
\end{eqnarray}

Although problems $\mathscr{P}$ and $\mathscr{D}$ are still nonconvex due to the nonconvex objective functions, the resulting smoothed $\ell_p$-minimization problems preserve the algorithmic
advantages, as will be presented in Section {\ref{em}}. {\rev{In particular, an iterative reweighted-$\ell_2$ algorithm will be developed in Section {\ref{em}} based on the principle of the MM algorithm to find a stationary point to the non-convex smoothed $\ell_p$-minimization problems $\mathscr{P}$ and $\mathscr{D}$.}}

\subsection{A Sparse Optimization Framework for Network Power Minimization with User Admission Control}
\begin{figure}[t]
  \centering
  \includegraphics[width=0.95\columnwidth]{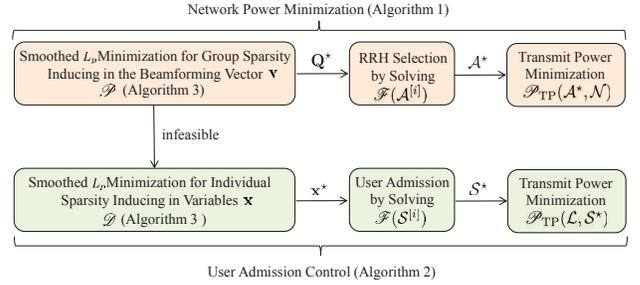}
 \caption{Sparse optimization for network power minimization and user admission control in multicast Cloud-RAN.}
 \label{framework}
\end{figure}
Denote the solutions of problems $\mathscr{P}$ and $\mathscr{D}$ as ${\bf{Q}}^{\star}$ and ${\bf{x}}^{\star}$, respectively.  Based on these induced (approximated) sparse solutions, we propose a sparse optimization framework for network power minimization  and user admission control in multicast Cloud-RAN. The main idea is illustrated in Fig. {\ref{framework}} and details will be explained in the following. In particular, Algorithm 3 will be
developed in Section {\ref{em}}, which yields a KKT point for problems $\mathscr{P}$ and $\mathscr{D}$.  

\subsubsection{Network Power Minimization} 
 If problem $\mathscr{P}$ is feasible, once obtaining its solution
${\bf{Q}}^{\star}$, we can extract the group sparsity structure information in the beamforming vector $\bf{v}$ based on the relation: $\|\tilde{\bf{v}}_l\|_2=\sqrt{\sum_{m=1}^{M}{\rm{Tr}}({\bf{C}}_{lm}{\bf{Q}}_m)}$, which will be zero if all the beamforming coefficients in $\tilde{\bf{v}}_l$ are zeros simultaneously. {\rev{By further incorporating system parameters to improve the performance \cite{Yuanming_TWC2014}, we adopt the following RRH ordering criteria to determine the priorities of the RRHs that should be switched off to minimize the network power consumption,    
\begin{eqnarray}
\label{ro}
\theta_{l}={\sqrt{{{\eta_l\kappa_l}\over{P_l^c}}}}\left({\sum\nolimits_{m=1}^{M}{\rm{Tr}}({\bf{C}}_{lm}{\bf{Q}}_m^{\star})}\right)^{1/2},
\forall l\in\mathcal{L},
\end{eqnarray}
where $\kappa_l=\sum_{k=1}^{K}\|{\bf{h}}_{kl}\|_2^2$ is the channel gain from the $l$-th RRH to all the MUs. The RRHs with a smaller parameter
$\theta_l$ will have a higher
priority to be switched off. Intuitively, the RRH with a lower channel power gain $\kappa_l$, lower drain inefficiency efficiency $\eta_l$, lower beamforming
gain $\|\tilde{\bf{v}}_l\|_2$, and higher relative fronthaul link power consumption $P_{l}^c$, should have a higher priority to be switched off.}}  

In this paper, we adopt a simple RRH selection procedure, i.e.,
the bi-section method, to switch off RRHs. This method was shown to provide good performance in \cite{Yuanming_TWC2014}. Specifically, based on the RRH ordering
rule in (\ref{ro}), we sort the coefficients in the ascending order: $\theta_{\pi_1}\le
\theta_{\pi_2}\le\dots\le\theta_{\pi_L}$ to determine the active RRHs. Let
$J_0$ be the maximum number of RRHs that can be switched off such that the
remaining RRHs can support the QoS requirements for all the MUs. To find $J_0$, in each bi-section search iteration, we need to solve  the following size-reduced convex feasibility problems based on the SDR technique,
\begin{eqnarray}
\label{fb1}
\mathscr{F}(\mathcal{A}^{[i]}): \mathop {\textrm{find}}&& {\bf{Q}}_1^{[i]},\dots, {\bf{Q}}_M^{[i]}\nonumber\\
\textrm{subject to}&&L_{k,m}(\{{\bf{Q}}_m^{[i]}\}_{m\in\mathcal{M}})\le 0, \forall k\in\mathcal{G}_m,
\nonumber\\
&& {\bf{Q}}_m^{[i]}\succeq {\bf{0}}, {\bf{Q}}_m^{[i]}\in\mathcal{Q}^{[i]}, \forall m\in\mathcal{M},
\end{eqnarray}
where ${\bf{Q}}_m^{[i]}\in\mathbb{C}^{(\sum_{l\in\mathcal{A}^{[i]}}N_l)\times
(\sum_{l\in\mathcal{A}^{[i]}}N_l)}$ with $\mathcal{A}^{[i]}=\{\pi_{i+1}, \dots, \pi_{L}\}$ as the active RRH set, ${\mathcal{Q}}^{[i]}$ represents the per-RRH transmit power constraints for the active RRHs in $\mathcal{A}^{[i]}$ and the QoS constraints are obtained after omitting the channel coefficients corresponding to the left-out RRHs. {\rev{If problem $\mathscr{F}(\mathcal{A}^{[i]})$ is feasible, it implies that a feasible solution exists to $\mathscr{F}(\mathcal{A}^{[J]})$ for all $J<i$. Likewise, if problem $\mathscr{F}(\mathcal{A}^{[i]})$ is infeasible, it implies that no feasible solution exists for any $J>i$. Therefore, determining the largest $J=J_0$ that results in a feasible solution to problem $\mathscr{F}(\mathcal{A}^{[J]})$ can be accomplished by solving no more than  $(1+\lceil\log(1+L)\rceil)$ such feasibility problems (\ref{fb1}) via bi-section search \cite{boyd2004convex}}}. {{Specifically, the set $\{0,1,\dots, L\}$ is guaranteed to contain
$J_0$, i.e., $J_0\in\{0,1,\dots, L\}$ at each step. In each iteration, the
set is divided in two sets, i.e., bisected, so the length of the set after
$k$ iterations is $2^{-k}(L+1)$ with $(L+1)$ as the length of the initial
set. It follows that exactly $(1+\lceil\log_2(1+L)\rceil)$ iterations are
required before the bi-section algorithm terminates. }} This procedure mainly reduces the relative fronthaul link power  consumption by switching off RRHs and the corresponding fronthaul links. 

Finally, denote the set of active RRHs as $\mathcal{A}^{\star}=\{\pi_{J_0+1},\dots, \pi_{L}\}$. To further reduce the network power consumption, we need to solve the following size-reduced
transmit power minimization problem with $|\mathcal{A}^{\star}|$ RRHs and $|\mathcal{N}|$
MUs based on the SDR technique,
\begin{eqnarray}
\label{tm}
\mathscr{P}_{\textrm{TP}}(\mathcal{A}^{\star}, \mathcal{N}):\mathop {\textrm{minimize}}_{{\bf{Q}}^{[J_0]}\in\mathcal{Q}^{[J_0]}}&&
\sum_{l\in\mathcal{A}^{\star}}\sum_{m=1}^{M}{1\over\eta_l}{\rm{Tr}}({\bf{C}}_{lm}{\bf{Q}}_m^{[J_0]})
 \nonumber\\
\textrm{subject to}&&L_{k,m}({\bf{Q}}^{[J_0]})\le 0, \forall k\in\mathcal{G}_m, 
\nonumber\\
&& {\bf{Q}}_m^{[J_0]}\succeq {\bf{0}}, \forall m\in\mathcal{M},
\end{eqnarray}
which is a semidefinite programming (SDP) problem and can be solved in polynomial time using the interior-point method.

The algorithm for solving the network power minimization problem is presented in Algorithm {\ref{algnp}}.

\begin{algorithm}
\label{algnp}
\caption{Network Power Minimization}
\textbf{Step 0:} Solve the group sparse inducing optimization
problem $\mathscr{P}$ (\ref{gsid}) using Algorithm 3 in Section {\ref{em}}.\\
\begin{enumerate}
\item {\textbf{If}} it is infeasible, {\textbf{go
to Algorithm 2}} for user admission control.
\item  {\textbf{If}} it is feasible, obtain the solutions ${\bf{Q}}_m^{\star}$'s,
calculate the ordering criterion  (\ref{ro}), and sort them
in
the ascending order: $\theta_{\pi_{1}}\le\dots\le\theta_{\pi_{L}}$, {\textbf{go
to Step 1}}.
\end{enumerate}
\textbf{Step 1:} Initialize $J_{\textrm{low}}=0$, $J_{\textrm{up}}=L$, $i=0$.\\
\textbf{Step 2:} Repeat
\begin{enumerate}
\item Set $i\leftarrow\lfloor{{J_{\textrm{low}}+J_{\textrm{up}}}\over{2}}\rfloor$.\\
\item Solve  problem $\mathscr{F}(\mathcal{A}^{[i]})$ (\ref{fb1}):
if it is infeasible, set $ J_{\textrm{up}}=i$; otherwise, set $J_{\textrm{low}}=i$.
\end{enumerate}
\textbf{Step 3:} Until $J_{\textrm{up}}-J_{\textrm{low}}=1$, obtain $J_{0}=J_{\textrm{low}}$
and obtain the optimal active RRH set $\mathcal{A}^{\star}=\{\pi_{J_0+1},\dots,
\pi_{L}\}$.\\
\textbf{Step 4:} Solve problem $\mathscr{P}_{\textrm{TP}}(\mathcal{A}^{\star},
\mathcal{N})$
(\ref{tm}) to obtain the multicast beamforming vectors for the
active RRHs.\\
\textbf{End}
\end{algorithm}

\subsubsection{User Admission Control}
When problem $\mathscr{P}$ is infeasible, we need to perform user admission control to maximize the user capacity. Specifically, let ${\bf{x}}^{\star}$ be the solution to the individual sparsity inducing optimization problem $\mathscr{D}$. Observe that ${{x}}_k$ represents the gap between the target SINR and the achievable SINR for MU $k$. We thus propose to admit the MUs with the smallest $x_i$'s \cite{Tan_JSAC2014, Tony_2015heterogeneous}. We order the coefficients in the descending
order: $x_{\pi_1}\ge x_{\pi_{2}}\ge\dots\ge x_{N}$. The bi-section search procedure will be adopted to find the maximum number of admitted MUs. Let $N_0$ be the minimum
number of MUs to be removed such that all the RRHs can support the QoS
requirements for all the remaining MUs. To determine the value of $N_0$, a sequence of the following convex sized-reduced feasibility problems need to be solved,
\begin{eqnarray}
\label{fb2}
\mathscr{F}(\mathcal{S}^{[i]}): \mathop {\textrm{find}}&& \{{\bf{Q}}_m\}_{m\in\mathcal{M}^{[i]}}\nonumber\\
\textrm{subject to}&&L_{k,m}(\{{\bf{Q}}_m\}_{m\in\mathcal{M}^{[i]}})\le 0, \forall k\in\mathcal{G}_m,
\nonumber\\
&& {\bf{Q}}_m\succeq {\bf{0}}, {\bf{Q}}_m\in\mathcal{Q}^{[i]}, \forall
m\in\mathcal{M}^{[i]},
\end{eqnarray}
where $\mathcal{S}^{[i]}=\{\pi_{i+1}, \dots, \pi_K\}$ denotes the set of admitted MUs, $\mathcal{M}^{[i]}=\{m:\mathcal{G}_m\cap\mathcal{S}^{[i]}\ne\emptyset\}$ is the set of multicast groups, and $\mathcal{Q}^{[i]}$ represents the per-RRH transmit power constraints with the served multicast groups $\mathcal{M}^{[i]}$. In this way, the  QoS constraints of the admitted MUs will be satisfied.

Finally, let $\mathcal{S}=\{\pi_{N_0+1},\dots,
\pi_{K}\}$ be the admitted MUs. We need to solve the same type of size-reduced transmit power minimization problem (\ref{tm}) with $|\mathcal{L}|$ RRHs  and $|\mathcal{S}^{\star}|$ MUs to find the multicast transmit beamforming vectors for all the admitted MUs. We denote this problem as $\mathscr{P}_{\textrm{TP}}(\mathcal{L}, \mathcal{S}^{\star})$. 

The proposed user admission control algorithm is presented in Algorithm {\ref{algua}}.

\begin{algorithm}
\label{algua}
\caption{User Admission Control}
\textbf{Step 0:} Solve the individual sparse inducing optimization
problem $\mathscr{D}$ (\ref{indsp}) using Algorithm 3 in Section {\ref{em}}. Obtain the solution ${\bf{x}}^{\star}$ and  sort the entries in the descending order: $x_{\pi_1}\ge\dots\ge x_{N}$, {\bf{go to Step 1}}.\\
\textbf{Step 1:} Initialize $N_{\textrm{low}}=0$, $N_{\textrm{up}}=K$, $i=0$.\\
\textbf{Step 2:} Repeat
\begin{enumerate}
\item Set $i\leftarrow\left\lfloor{{N_{\textrm{low}}+N_{\textrm{up}}}\over{2}}\right\rfloor$.\\
\item Solve  problem $\mathscr{F}(\mathcal{S}^{[i]})$ (\ref{fb2}):
if it is feasible, set $N_{\textrm{up}}=i$; otherwise, set $N_{\textrm{low}}=i$.
\end{enumerate}
\textbf{Step 3:} Until $N_{\textrm{up}}-N_{\textrm{low}}=1$, obtain $N_{0}=N_{\textrm{up}}$
and obtain the  admitted MU set $\mathcal{S}^{\star}=\{\pi_{N_0+1},\dots,
\pi_{K}\}$.\\
\textbf{Step 4:} Solve problem $\mathscr{P}_{\textrm{TP}}(\mathcal{L},
\mathcal{S}^{\star})$
(\ref{tm}) to obtain the multicast beamforming vectors for the
admitted MUs.\\
\textbf{End}
\end{algorithm}

\begin{remark}[Rank-One Approximation  After SDR]
The solutions for the SDR based optimization problems $\mathscr{P}$, $\mathscr{D}$, $\mathscr{F}(\mathcal{A}^{[i]})$, $\mathscr{F}(\mathcal{S}^{[i]})$ and $\mathscr{P}_{\textrm{TP}}(\mathcal{A},\mathcal{S})$ may not be rank-one. 
If the rank-one solutions are failed to be obtained,
the Gaussian randomization method \cite{Luo_2008quality} will be employed to obtain the  feasible
rank-one approximate solution. {\rev{Specifically, the candidate multicast beamforming vectors are generated from the
solution of the SDR problems, and one is picked yielding a feasible solution to the original problem
with the minimum value of the objective function. The feasibility for the original problem is guaranteed
by solving a sequence of multigroup multicast power control problems with the fixed beamforming directions via
linear programming \cite{Luo_2008quality}. Please refer to \cite[Section IV]{Luo_2008quality} for more details on the Gaussian randomization method and we adopt this method in our simulations to find approximate rank-one feasible solutions.}} While the optimality of this randomization method for general problems remains unknown, it has been widely applied and shown to provide good performance \cite{Luo_2006transmitmulticasting,Mehanna_SP2013}.

\end{remark}  

\subsubsection{Complexity Analysis and Discussions}
\label{comad}
{\rev{To implement Algorithm {\ref{algnp}} and Algorithm {\ref{algua}}, a sequence of SDP optimization or feasibility problems (e.g., $\mathscr{P}$, $\mathscr{D}$,
$\mathscr{F}(\mathcal{A}^{[i]})$, $\mathscr{F}(\mathcal{S}^{[i]})$ and $\mathscr{P}_{\textrm{TP}}(\mathcal{A},\mathcal{S})$) need to be solved. In particular, to find the active RRH set $\mathcal{A}^{\star}$ and admitted MU set $\mathcal{S}^{\star}$, we need to solve no more than $(1+\lceil\log(1+L)\rceil)$ and $(1+\lceil\log(1+K)\rceil)$ SDP feasibility problems $\mathscr{F}(\mathcal{A}^{[i]})$ and $\mathscr{F}(\mathcal{S}^{[i]})$, respectively. In addition, to solve the SDP problem $\mathscr{P}_{\textrm{TP}}(\mathcal{L},
\mathcal{N})$
(\ref{tm}) with $M$ matrix optimization variables of size $N\times N$ and $(K+L)$ linear constraints, the interior-point method \cite{boyd2004convex} will take $\mathcal{O}(\sqrt{MN}\log(1/\epsilon))$ iterations and cost $\mathcal{O}(M^3N^6+(K+L)MN^2)$ floating point operations to achieve an optimal solution with accuracy $\epsilon>0$. Therefore, this makes the proposed network power minimization and user admission algorithms difficult to scale to large problem sizes with a large number of RRHs and/or MUs. To further improve the computational efficiency of the proposed SDP based algorithms, one promising approach is to apply the alternating direction method of multipliers (ADMM) algorithm \cite{boyd2011distributed} by leveraging parallelism in the cloud computing environment in the BBU pool \cite{Yuanming_LargeSOCP2014}. This is, however, an on-going research topic, and we will leave it as our future work.}}

\section{Iterative Reweighted-$\ell_2$ Algorithm for  Smoothed $\ell_p$-Minimization}
\label{em}
In this section, we first develop an iterative reweighted-$\ell_2$ algorithm to solve a general  non-convex smoothed $\ell_p$-minimization problem based on the principle of the MM algorithm. We then present how to apply this algorithm to solve the problems $\mathscr{P}$ and $\mathscr{D}$ to induce sparsity structures in the solutions, thereby guiding the  RRH selection and user admission.

\subsection{Iterative Reweighted-$\ell_2$ Algorithm}

Consider  the following smoothed
$\ell_p$-minimization problem, 
\begin{eqnarray}
\label{lps}
\mathscr{P}_{\textrm{sm}}(\epsilon): \mathop {\textrm{minimize}}_{{\bs{z}}\in\mathcal{C}}&& f_{p}({\bs{z}}; \epsilon):=\sum_{i=1}^{m}(z_i^2+\epsilon^2)^{p/2},
\end{eqnarray}
where $\mathcal{C}$ is an arbitrary convex set, ${\bs{z}}\in\mathbb{R}^{m}$ and $\epsilon>0$ is some fixed regularizing parameter. In the following, we first prove that the optimal solution of the smoothed $\ell_p$-minimization problem $\mathscr{P}_{\textrm{sm}}(\epsilon)$ is also optimal for the original non-smooth $\ell_p$-minimization problem (i.e., $\mathscr{P}_{\textrm{sm}}(0)$) when $\epsilon$ is small. We then demonstrate the algorithmic advantages  of the smoothness in the procedure of developing the iterative reweighted-$\ell_2$ algorithm.  

\subsubsection{Optimality of Smoothing the $\ell_p$-Norm}
The set of KKT {\rev{points}} of problem $\mathscr{P}_{\textrm{sm}}(\epsilon)$ is given as
\begin{eqnarray}
\label{kktd}
\Omega({\epsilon})=\{{\bs{z}}\in\mathcal{C}: 0\in \nabla_{\bs{z}}f_p({\bs{z}}; \epsilon)+\mathcal{N}_{\mathcal{C}}({\bs{z}})\},
\end{eqnarray}
where $\mathcal{N}_{\mathcal{C}}({\bs{z}})$ is the normal cone of a convex set $\mathcal{C}$ at point ${\bs{z}}$ consisting of the outward normals to all hyperplanes that support $\mathcal{C}$ at ${\bs{z}}$, i.e., 
\begin{eqnarray}
\mathcal{N}_{\mathcal{C}}({\bs{z}}):=\{{\bs{s}}: \langle {\bs{s}}, {\bf{x}}-{\bs{z}}\rangle\le 0, \forall {\bf{x}}\in\mathcal{C}\}.
\end{eqnarray}
Define the deviation of a given set $\mathcal{Z}_1$ from another set $\mathcal{Z}_2$ as \cite{shapiro2009lectures},
\begin{eqnarray}
\mathbb{D}(\mathcal{Z}_1, \mathcal{Z}_2)=\sup_{z_1\in\mathcal{Z}_1}\left(\inf_{z_2\in\mathcal{Z}_2}\|z_1-z_2\|\right).
\end{eqnarray} 
We then have the following  theorem on the relationship between the smoothed
$\ell_p$-minimization problem $\mathscr{P}_{\textrm{sm}}(\epsilon)$ and the original non-smooth $\ell_p$-minimization problem $\mathscr{P}_{\textrm{sm}}(0)$. 

\begin{theorem}
\label{kkt}
Let $\Omega_{\epsilon}$ be the set of KKT points of problem $\mathscr{P}_{\textrm{sm}}(\epsilon)$. Then, we have
\begin{eqnarray}
\label{kkt1}
\lim_{\epsilon\searrow {0}}\mathbb{D}(\Omega({\epsilon}), \Omega(0))=0.
\end{eqnarray}
\begin{IEEEproof}
Please refer to Appendix {\ref{kktap}} for details.  
\end{IEEEproof}
\end{theorem}

This theorem indicates that any limit of the sequence of KKT {\rev{points}} of problem $\mathscr{P}_{\textrm{sm}}(\epsilon)$ is a KKT pair of problem $\mathscr{P}_{\textrm{sm}}(0)$ when $\epsilon$ is small enough. That is, at least a local optimal solution can be achieved. In the sequel, we will focus on finding a KKT point of problem $\mathscr{P}_{\textrm{sm}}({\epsilon})$ with a small $\epsilon$, yielding good approximations to the KKT points of the $\ell_p$-minimization problem $\mathscr{P}_{\textrm{sm}}(0)$ to induce sparsity in the solutions.

\subsubsection{The MM Algorithm for the Smoothed $\ell_p$-Minimization}
With the established asymptotic optimality, we then leverage the principle of  the MM algorithm
to solve problem (\ref{lps}). Basically, this algorithm generates the iterates $\{{\bs{z}}_{n}\}_{n=1}^{\infty}$ by successively minimizing upper bounds $Q({\bs{z}}; {\bs{z}}^{[n]})$ of the objective function $f_p({\bs{z}};\epsilon)$. The quality of the upper bounds will control the convergence (rate) and optimality of the resulting algorithms. Inspired by the results in the expectation-maximization (EM) algorithm \cite{Lange_1993normal,Ba_2014convergenceTSP}, we adopt the upper bounds in the following proposition to approximate the smoothed $\ell_p$-norm.    

\begin{proposition}
\label{pro1}
Given the iterate ${\bs{z}}^{[n]}$ at the $n$-th iteration, an upper bound for the objective function of the smoothed $\ell_p$-norm $f_{p}({\bs{z}};\epsilon)$ can be constructed as follows,
\begin{eqnarray}
\label{qf}
Q({\bs{z}}; {\bs{\omega}}^{[n]}):=\sum_{i=1}^{m}\omega_i^{[n]}z_i^2,
\end{eqnarray}
where 
\begin{eqnarray}
\label{wup}
{\omega}_i^{[n]}={p\over{2}}{{\left[\left(z_i^{[n]}\right)^2+\epsilon^2\right]^{{p\over{2}}-1}}},
\forall i=1,\dots, m.
\end{eqnarray}
From the weights given in (\ref{wup}), it is clear that, by adding the regularizer parameter $\epsilon>0$, we can avoid yielding infinite values when some $z_i$'s become zeros in the iterations. 

\begin{IEEEproof}
Define the approximation error as
\begin{eqnarray}
\label{error}
\!\!\!f_{p}({\bs{z}};\epsilon)-Q({\bs{z}}; {\bs{\omega}}^{[n]})=\sum_{i=1}^{m}[\kappa(z_i^2)-\kappa'((z_i^{[n]})^2)z_i^2],
\end{eqnarray}
where $\kappa(s)=(s+\epsilon^2)^{p/2}$ with $s\ge 0$. The sound property of the $Q$-function (\ref{qf}) is that the approximation error (\ref{error}) attains its maximum at ${\bs{z}}={\bs{z}}^{[n]}$. In particular, we only need to prove that the function $g(s)=\kappa(s)-\kappa'(s^{[n]})s$ with $s\ge 0$ attains the maximum at $s=s^{[n]}$. This is true based on the facts that $g'(s^{[n]})=0$ and $\kappa(s)$ is strictly concave.  
\end{IEEEproof}
\end{proposition}

Let ${\bs{z}}^{[n+1]}$ be the minimizer of the upper bound function $Q({\bs{z}}; {\bs{\omega}}^{[n]})$ at the $n$-th iteration, i.e.,
\begin{eqnarray}
\label{qm}
{\bs{z}}^{[n+1]}:=\arg\min_{{\bs{z}}\in\mathcal{C}}~Q({\bs{z}}; {\bs{\omega}}^{[n]}).
\end{eqnarray}
Based on Proposition {\ref{pro1}} and (\ref{qm}), we have  
\begin{eqnarray}
\label{ub1}
f_p({\bs{z}}^{[n+1]};\epsilon)&=&Q({\bs{z}}^{[n+1]};{\bs{\omega}}^{[n]})+f_p({\bs{z}}^{[n+1]};\epsilon)\nonumber\\
&&-Q({\bs{z}}^{[n+1]};
{\bs{\omega}}^{[n]})\nonumber\\
&\le& Q({\bs{z}}^{[n+1]};{\bs{\omega}}^{[n]})+f_p({\bs{z}}^{[n]};\epsilon)-Q({\bs{z}}^{[n]};
{\bs{\omega}}^{[n]})\nonumber\\
&\le& Q({\bs{z}}^{[n]}; {\bs{\omega}}^{[n]})+f_p({\bs{z}}^{[n]};\epsilon)-Q({\bs{z}}^{[n]};
{\bs{\omega}}^{[n]})\nonumber\\
&=&f_p({\bs{z}}^{[n]};\epsilon),
\end{eqnarray}
where the first inequality is based on the fact that function $(f_{p}({\bs{z}};\epsilon)-Q({\bs{z}}; {\bs{\omega}}^{[n]}))$ attains its maximum at ${\bs{z}}={\bs{z}}^{[n]}$, and the second inequality follows from (\ref{qm}).
Therefore, minimizing the upper bound, i.e., the $Q$-function in (\ref{qf}), can reduce the objective function $f_{p}({\bs{z}};\epsilon)$ successively. 

\begin{remark}
In the context of the EM algorithm \cite{Dempster_1977maximum} for computing the maximum likelihood estimator of latent
variable models, the functions $-f_{p}({\bs{z}};\epsilon)$ and $-Q({\bs{z}}; {\bs{\omega}}^{[n]})$ can be regarded as the log-likelihood and comparison functions (i.e., the lower bound of the log-likelihood), respectively \cite{Lange_1993normal}.
\end{remark} 

The MM algorithm for the smoothed $\ell_p$-minimization problem is presented in Algorithm {\ref{irm}}. 
\begin{algorithm}
\label{irm}
\caption{Iterative Reweighted-$\ell_2$ Algorithm}
{\textbf{input}}: Initialize ${\bs{\omega}}^{[0]}=(1, \dots, 1)$;
$I$ (the maximum number of iterations)\\
 Repeat
 
~~1) Solve problem 
\begin{eqnarray}
\label{irls}
{\bs{z}}^{[n+1]}:=\arg\min_{{\bs{z}}\in\mathcal{C}} \sum_{i=1}^{m}\omega_i^{[n]}z_i^2.
\end{eqnarray}
If it is feasible, {\textbf{go to}} 2); otherwise, {\textbf{stop}} and return
{\textbf{output
2}}.

~~2) Update the weights as
\begin{eqnarray}
\label{w22}
{\omega}_i^{[n+1]}={p\over{2}}{{\left[\left(z_i^{[n+1]}\right)^2+\epsilon^2\right]^{{p\over{2}}-1}}},
\forall i=1,\dots, m.
\end{eqnarray}

Until convergence or attain the  maximum iterations and return {\textbf{output
1}}.\\
 {\textbf{output 1}}: ${\bs{z}}^{\star}$; {\textbf{output
2}}: Infeasible.
\label{gsbfal} 
\end{algorithm} 

The convergence of the iterates $\{{\bs{z}}^{[n]}\}_{n=1}^{\infty}$ (\ref{irls}) is presented in the following theorem. 
\begin{theorem}
\label{thecon}
Let $\{{\bs{z}}^{[n]}\}_{n=1}^{\infty}$ be the sequence generated by the iterative reweighted-$\ell_2$ algorithm (\ref{irls}). Then, every limit point $\bar{\bs{z}}$ of $\{{\bs{z}}^{[n]}\}_{n=1}^{\infty}$ has the following properties
\begin{enumerate}
\item $\bar{\bs{z}}$ is a KKT point of problem $\mathscr{P}_{\textrm{sm}}(\epsilon)$ (\ref{lps});
\item $f_{p}({\bs{z}}^{[n]}; \epsilon)$ converges monotonically to $f_{p}({\bs{z}}^{\star}; \epsilon)$
for some KKT point ${\bs{z}}^{\star}$.
\end{enumerate}

\begin{IEEEproof}
Please refer to Appendix {\ref{converge}} for details. 
\end{IEEEproof}

\end{theorem}

{\rev{As noted in \cite{bertsekas1999nonlinear}, without the convexity of $f_p({\bs{z}};\epsilon)$, the KKT point may be a local minimum or other point (e.g., a saddle point). We also refer to these points as stationary points \cite[Page 194]{bertsekas1999nonlinear}.}}

\begin{remark}
The algorithm consisting of the iterate (\ref{irls}) accompanied with weights (\ref{w22}) is known as the \emph{iterative reweighted least squares} \cite{Daubechies_2010iteratively,Dempster_1977maximum,Yin_2008iteratively} in the fields of statistics, machine learning and compressive sensing. In particular, with a simple constraint $\mathcal{C}$, the iterates often yield closed-forms with better computational efficiency. For instance, for the noiseless compressive sensing problem \cite{Daubechies_2010iteratively}, the iterates have closed-form solutions \cite[(1.9)]{Daubechies_2010iteratively}. {\rev{Therefore, this method has a higher computational efficiency compared with the conventional $\ell_1$-minimization approach for compressive sensing \cite{Tao_IT06}, wherein a linear programming problem needs to be solved via algorithms such as interior-point or barrier methods. Furthermore, empirically, it was observed that the iterative reweighted least squares method can improve the signal recovery capability by enhancing the sparsity for compressive sensing over the $\ell_1$-minimization method \cite{Daubechies_2010iteratively, Yin_2008iteratively}.}}
\end{remark}

In contrast to the existing works on the iterative reweighted least squares methods, we provide a new perspective
to develop the iterative reweighted-$\ell_2$ algorithm to solve the smoothed  $\ell_p$-minimization problem with convergence guarantees based on the principle
of the MM algorithm. Furthermore, the main motivation and advantages for developing the
iterates (\ref{irls}) is to induce the quadratic forms in the objective function in problem  $\mathscr{P}$ (\ref{gsid}) to make it compliant with the SDR technique, thereby inducing the group sparsity structure in the multicast beamforming vectors via convex programming.

\begin{remark}
{\rev{The advantages of the iterative reweighted-$\ell_2$ algorithm include the capability of  enhancing the sparsity, as well as inducing the quadratic forms for the multicast beamforming vectors by  inducing the quadratic formulations (\ref{irls}). {\rev{Note that the reweighted $\ell_1$-minimization algorithm in \cite{Boyd_2008enhancing} can also induce the quadratic forms in the beamforming vectors $\tilde{\bf{v}}_l$'s by rewriting the indicator function (\ref{fh1}) as the $\ell_0$-norm of the squared $\ell_2$-norm of the vectors $\tilde{\bf{v}}_l$'s, i.e., $\|\tilde{\bf{v}}_l\|_2^2$}}. Furthermore,  the key ideas of the convergence proof of the iterative reweighted-$\ell_2$ algorithm (i.e., Algorithm {\ref{irm}}), by leveraging the EM theory to establish upper bounds in the iterates of the MM algorithm, should be useful for other iterative algorithms, e.g., \cite{Wei_Globcom13sparse}.}}
\end{remark}

\subsection{Sparsity Inducing for RRH Selection and User Admission} 
In this subsection, we demonstrate how to apply the developed iterative reweighted-$\ell_2$ algorithm to solve the {\rev{nonconvex}} sparse optimization problems $\mathscr{P}$ and $\mathscr{D}$ for RRH selection and user admission, respectively. {\rev{In this way, we can find a KKT point for the nonconvex smoothed $\ell_p$-minimization problems $\mathscr{P}$ and $\mathscr{D}$ with convex constraints}}. Specifically, let $\Omega:=\{{\bf{Q}}\in\mathcal{Q}: L_{k,m}({\bf{Q}})\le 0, {\bf{Q}}_m\succeq {\bf{0}}, \forall k\in\mathcal{G}_m, m\in\mathcal{M}\}$ be the set of constraints in problem $\mathscr{P}$. The iterative reweighted-$\ell_2$ algorithm for problem $\mathscr{P}$
generates a sequence $\{{\bf{Q}}^{[n]}\}_{n=1}^{\infty}$ as follows:
\begin{eqnarray}
\label{sdp1}
\mathscr{P}_{\textrm{SDP}}^{[n]}: \mathop {\textrm{minimize}}_{{\bf{Q}}\in\Omega}&&
\sum_{l=1}^{L}\omega_l^{[n]}\left(\sum_{m=1}^{M}{\rm{Tr}}({\bf{C}}_{lm}{\bf{Q}}_m)\right),
\end{eqnarray}
with the weights as 
\begin{eqnarray}
\label{w1}
\!\!\omega_l^{[n]}={\rho_l p\over{2}}\left[\sum\nolimits_{m=1}^{M}{\rm{Tr}}\left({\bf{C}}_{lm}{\bf{Q}}_m^{[n]}\right)+\epsilon^{2}\right]^{{p\over{2}}-1}, \forall l\in\mathcal{L}.
\end{eqnarray}

Applying Algorithm {\ref{irm}} to problem $\mathscr{D}$ is straightforward.

\section{Simulation Results}
\label{simres}
In this section, we will simulate the proposed algorithms based on the iterative reweighted-$\ell_2$ algorithm (IR2A) for network power minimization and user admission control in multicast Cloud-RAN. We set the parameters as follows: $P_l=1 W ,\forall l$; $P_l^c=[5.6+l-1]W, \forall l$; $\eta_l=1/4, \forall l$; $\sigma_k=1, \forall k$. Denote the channel propagation from the $l$-th RRH and the $k$-th MU as ${\bf{h}}_{kl}={D}_{kl}{\bf{g}}_{kl}$ with ${D_{kl}}$ as the large-scale fading coefficients and ${\bf{g}}_{kl}\in\mathcal{CN}({\bf{0}}, {\bf{I}})$ as the small-scale fading coefficients. {\rev{We set $\epsilon=10^{-3}$ in the iterative reweighted-$\ell_2$ algorithm and the algorithm will terminate if either the number of iterations exceeds 30 or the difference between the objective values of consecutive iterations is less than $10^{-3}$. The pre-determined number of randomization is set to be 50 in the Gaussian randomization method.}}

\subsection{Convergence of the Iterative Reweighted-$\ell_2$ Algorithm}
\label{simconv}
\begin{figure}[!t]
\centering
  \includegraphics[width=0.95\columnwidth]{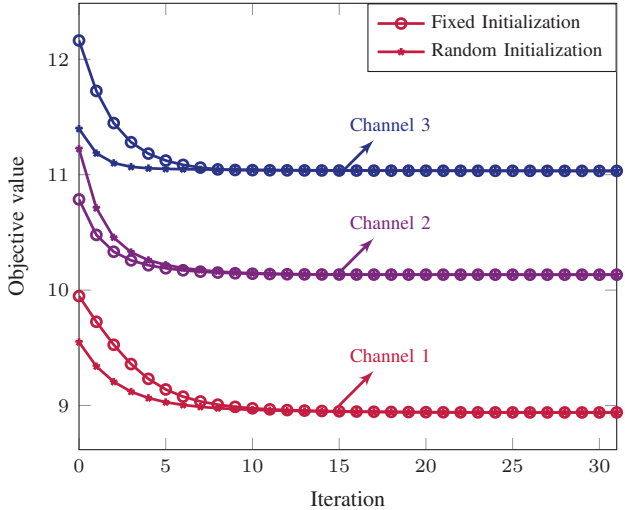}
\caption{Convergence of the iterative reweighted-$\ell_2$ algorithm for the
smoothed $\ell_p$-minimization problem $\mathscr{P}$ with different channel
realizations and initial points.}
\label{convergefig}
\end{figure}

Consider a network with $L=6$ 2-antenna RRHs and 2 multicast groups with 2 single-antenna MUs in each group. The channels are spatially uncorrelated. For each MU $k$, we set $D_{lk}=1, \forall l\in\Omega_1$ with $|\Omega_1|=2$; $D_{lk}=0.7, \forall l\in\Omega_2$ with $|\Omega_2|=2$; $D_{lk}=0.5, \forall l\in\Omega_3$ with $|\Omega_3|=2$. All the sets $\Omega_i$'s are uniformly drawn from the RRH set $\mathcal{L}=\{1,\dots, L\}$ with $\cup\Omega_i=\Omega$. {\rev{Fig. {\ref{convergefig}} illustrates the convergence of the iterative reweighted-$\ell_2$ algorithm for the smoothed $\ell_p$-minimization problem $\mathscr{P}$ with different initial points and different channel realizations. Specifically, we set $p=1$. The three
different channel realizations are generated uniformly and independently.
For each channel realization, we simulate Algorithm {\ref{irm}} with the
fixed initial point as ${\bs{\omega}}^{[0]}=(1, \dots, 1)$ and randomly generated initial point ${\bs{\omega}}^{[0]}$.
The random initialization instances for the three channel realizations are given as ${\bs{\omega}}_{{\rm{ch1}}}^{[0]}=[0.1389,0.2028,0.1987,0.6038,0.2722,0.1988]$,
${\bs{\omega}}_{{\rm{ch2}}}^{[0]}=[0.5657,0.7165,0.5113,0.7764,0.4893,0.1859]$,
${\bs{\omega}}_{{\rm{ch3}}}^{[0]}=[0.4093,0.4635,0.6109,0.0712,0.3143,0.6084]$, respectively. In particular, this figure shows that the sequence of the objective functions converges monotonically, which confirms the convergence results in Theorem {\ref{thecon}}. In addition, it illustrates the robustness of the convergence of the reweighted-$\ell_2$ algorithm with different initial points and different problem parameters.}} Furthermore, it also demonstrates the fast convergence rate of the proposed algorithm in the simulated setting. Empirically, the iterative reweighted-$\ell_2$ algorithm converges in 20 iterations on average in all the simulated channels in this paper.    

\subsection{Network Power Minimization}
 
\begin{figure}
\centering
  \includegraphics[width=0.95\columnwidth]{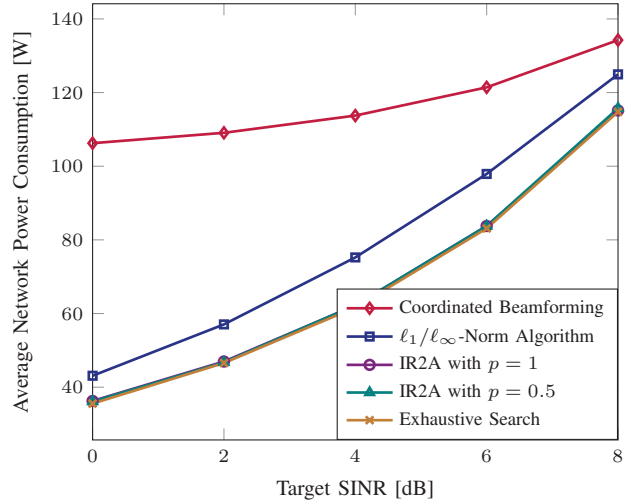}
\caption{Average network power consumption versus target SINR with different
algorithms.}
\label{npfig}
\end{figure}
 
 \begin{table}[!t]
\renewcommand{\arraystretch}{1.3}
\caption{The Average Number of Active RRHs with Different Algorithms}
\label{sparsity1}
\centering
\begin{tabular}{l|c|c|c|c|c}
\hline
\tabincell{c}{{Target SINR [dB]}} & 0 & 2 &
4 & 6 & 8 \\
\hline
Coordinated Beamforming & 12.00 & 12.00 & 12.00 & 12.00 & 12.00  \\
\hline
\tabincell{c}{$\ell_1/\ell_\infty$-Norm  Algorithm} & 4.39 & 5.43 & 6.70
&
8.35 & 10.35 \\
\hline
\tabincell{c}{IR2A  with $p=1$} & 4.74 & 5.43 & 6.65 &
8.22 & 10.13  \\
\hline
\tabincell{c}{IR2A  with $p=0.5$} & 4.61 & 5.30 & 6.65 &
8.22 & 10.17  \\
\hline
\tabincell{c}{Exhaustive Search} & 4.43 & 5.30 & 6.39 & 8.04 & 10.09\\
\hline
\end{tabular}
\label{rrh_num}
\end{table} 
 
 \begin{table}[!t]
\renewcommand{\arraystretch}{1.3}
\caption{The Average Relative Fronthaul Links Power Consumption with Different
Algorithms }
\label{sparsity1}
\centering
\begin{tabular}{l|c|c|c|c|c}
\hline
\tabincell{c}{{Target SINR [dB]}} & 0 & 2 &
4 & 6 & 8 \\
\hline
Coordinated Beamforming & 102.0 & 102.0 & 102.0 & 102.0 & 102.0  \\
\hline
\tabincell{c}{$\ell_1/\ell_\infty$-Norm  Algorithm} & 30.65 & 40.52 & 53.74
&
69.70 & 87.48 \\
\hline
\tabincell{c}{IR2A with $p=1$} & 24.57 & 29.91 & 40.26
&
56.30 & 78.04 \\
\hline
\tabincell{c}{IR2A with $p=0.5$} & 24.13 & 29.39 & 40.43
&
56.30 & 78.87 \\
\hline
\tabincell{c}{Exhaustive Search} & 22.96 & 29.65 & 39.00 &
54.78 & 77.22  \\
\hline
\end{tabular}
\label{fronthaulpower}
\end{table}

\begin{table}[!t]
\renewcommand{\arraystretch}{1.3}
\caption{The Average Total Transmit Power Consumption with Different Algorithms}
\label{sparsity1}
\centering
\begin{tabular}{l|c|c|c|c|c}
\hline
\tabincell{c}{{Target SINR [dB]}} & 0 & 2 &
4 & 6 & 8 \\
\hline
Coordinated Beamforming & 4.26 & 7.07 & 11.74 & 19.42 & 32.24  \\
\hline
\tabincell{c}{$\ell_1/\ell_\infty$-Norm  Algorithm} & 12.45 & 16.53 & 21.49
&
28.20 & 37.46 \\
\hline
\tabincell{c}{IR2A with $p=1$} & 11.69 & 17.05 & 21.94 &
27.46 & 37.10  \\
\hline
\tabincell{c}{IR2A with $p=0.5$} & 11.97 & 17.36 & 21.79 &
27.46 & 36.85  \\
\hline
\tabincell{c}{Exhaustive Search} & 12.59 & 16.91 & 22.62 &
28.33 & 37.53  \\
\hline
\end{tabular}
\label{transmitpower}
\end{table}

Consider a network with $L=12$ 2-antenna RRHs and 5 multicast groups with 2 single-antenna MUs in each group. The channels are spatially uncorrelated.
For each MU $k$, we set $D_{lk}=1, \forall l\in\Omega_1$ with $|\Omega_1|=4$;
$D_{lk}=0.7, \forall l\in\Omega_2$ with $|\Omega_2|=4$; $D_{lk}=0.5, \forall
l\in\Omega_3$ with $|\Omega_3|=4$. All the sets $\Omega_i$'s are uniformly
drawn from the RRH set $\mathcal{L}=\{1,\dots, L\}$ with $\cup\Omega_i=\Omega$. The proposed iterative reweighted-$\ell_2$  algorithm is compared with the reweighted $\ell_1/\ell_\infty$-norm based algorithm \cite{Mehanna_SP2013}, in which, the objective function in problem $\mathscr{P}$ is replaced by $\sum_{l_1=1}^{L}\sum_{l_2}^{L}\max_{m}\max_{n_{l_1}}\max_{n_{l_2}}|{\bf{Q}}_{m}(n_{l_1}, n_{l_2})|$ with ${\bf{Q}}_{m}(i,j)$ as the $(i,j)$-th entry in ${\bf{Q}}_m$. And the RRHs with  smaller beamforming beamforming coefficients (measured by $\ell_\infty$-norm) have higher priorities to be switched off. Fig. {\ref{npfig}} demonstrates the average network power consumption with different target SINRs using different algorithms. Each point of the simulation results is averaged over 50 randomly generated channel realizations for which problem $\mathscr{P}$ is feasible. From this figure, we can see that the proposed iterative reweighted-$\ell_2$  algorithm can achieve near-optimal performance compared with the exhaustive search algorithm ({\rev{i.e., solving problem (\ref{pm}) with convexified QoS constraints based on the SDR technique}}). It yields lower network power consumption compared with the existing $\ell_1/\ell_\infty$-norm based algorithm \cite{Mehanna_SP2013}, while the coordinated multicast beamforming algorithm \cite{WeiYu_WC10, Luo_2006transmitmulticasting} with all the RRHs active has the highest network power consumption. {\rev{Note that the number of SDP problems $\mathscr{F}(\mathcal{A}^{[i]})$ (\ref{fb1}) needed to be solved grows \emph{logarithmically} with $L$ for the proposed network power minimization algorithm and the previous $\ell_1/\ell_\infty$-norm based algorithm \cite{Mehanna_SP2013}. But the number of SDP problems (\ref{fb1}) to be solved for exhaustive search grows exponentially with $L$. Although the coordinated beamforming algorithm has the lowest computational complexity by only solving the total transmit
power minimization problem $\mathscr{P}_{\textrm{TP}}(\mathcal{L}, \mathcal{N})$
(\ref{tm}) with all the RRHs active, it yields the highest network power consumption. Our proposed algorithm thus achieves a good trade-off between computational complexity and performance.}}

Tables {\ref{rrh_num}}-{\ref{transmitpower}} show the corresponding average  number of active RRHs, average fronthaul network power consumption and average total transmit power consumption, respectively. Specifically, Table {\ref{rrh_num}} confirms the existence of the group sparsity pattern in the aggregative multicast beamforming vector. That is, the switched off RRHs indicate that all the beamforming coefficients at the RRHs are set to be zeros simultaneously. It also shows that the proposed iterative reweighted-$\ell_2$  algorithm has the capability of enhancing sparsity, thereby switching off more RRHs compared with the $\ell_1/\ell_\infty$-norm based algorithm except for very low SNR. Table {\ref{fronthaulpower}} shows that the proposed iterative reweighted-$\ell_2$  algorithm can achieve much lower fronthaul network power consumption and attain similar values with exhaustive search. With all the RRHs being active, the coordinated multicast beamforming algorithm yields the highest fronthaul network power consumption and the lowest transmit power consumption as shown in Table {\ref{fronthaulpower}} and Table {\ref{transmitpower}}, respectively. This indicates that it is critical to take the relative fronthaul link power consumption into consideration when designing a green Cloud-RAN. {\rev{Although the exhaustive search may yield higher transmit power consumption in some scenarios as shown in Table {\ref{transmitpower}}, overall it achieves the lowest network power consumption as much lower relative fronthaul network power consumption can be attained as indicated in Table {\ref{fronthaulpower}}}}. Furthermore, we can see that, in the simulated settings, different values of $p$ in the proposed iterative reweighted-$\ell_2$  algorithm yield similar performance, while all achieve near-optimal performance compared with the exhaustive search.           

\subsection{User Admission Control}
\begin{figure}
  \centering
  \includegraphics[width=0.95\columnwidth]{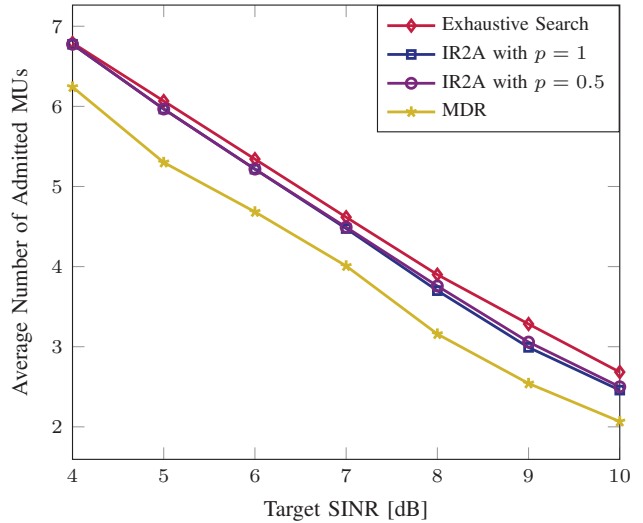}
\caption{Average number of admitted MUs versus target SINR with different algorithms.}
\label{userfig}
\end{figure}

Consider a network with $L=6$ 2-antenna RRHs and 4 multicast groups with
2 single-antenna MUs in each group. The channel model is the same as Section {\ref{simconv}}. The proposed iterative reweighted-$\ell_2$ algorithm based user admission control algorithm  is compared with the existing convex relaxation approach (i.e.. the multicast membership deflation by relaxation (MDR) \cite{Luo_TSP2009efficient}) and the exhaustive search. The simulation results are illustrated in Fig.{\ref{userfig}} for the average number of admitted MUs. Each point of the simulation results is averaged over 200 randomly generated channel realizations for which problem
$\mathscr{P}$ is infeasible.  From Fig. {\ref{userfig}}, we can see that the proposed iterative reweighted-$\ell_2$ algorithm outperforms the existing MDR approach \cite{Luo_TSP2009efficient}. In particular, the performance of the iterative reweighted-$\ell_2$ algorithm is almost the same with different values of $p$ and achieves near-optimal performance compared to the exhaustive search 
{\rev{via solving problem
(\ref{ua}) with convexified QoS constraints based on the SDR technique.}} {\rev{Although all the simulated results demonstrate that the performances
are robust to the parameter $p$, it is very interesting to theoretically identify the typical
scenarios, where smaller values of
$p$ will yield much better performances.}}

\section{Conclusions and Future Works}
\label{confu}
This paper developed a sparse optimization framework for  network power minimization and user admission in green Cloud-RAN with multicast beamforming. A smoothed $\ell_p$-minimization method was proposed to induce the sparsity structures in the solutions, thereby guiding the RRH selection and user admission. This approach has the advantages in terms of promoting sparsity and assisting algorithmic design by introducing the quadratic forms in the group sparse inducing penalty. {\rev{In particular, by leveraging the  MM algorithm and the SDR technique, we developed an iterative reweighted-$\ell_2$ algorithm with convergence and optimality guarantees (i.e., KKT points) to solve the resulting smoothed $\ell_p$-minimization problems $\mathscr{P}$ and $\mathscr{D}$ with convex constraints}}. The effectiveness of the proposed algorithms was demonstrated via simulations for  network power minimization and user admission control.       

Several future directions are listed as follows:
\begin{itemize}
\item Although the proposed methods only need to solve a sequence of convex optimization problems, the complexity of solving the large-scale SDP problem using the interior-point method is prohibitively high. One promising option is to use the first-order method (e.g., the operator splitting method \cite{Boyd_arXiv2013, Yuanming_LargeSOCP2014}) by leveraging the parallel computing platform in the BBU pool \cite{Yuanming_WCMLargeCVX,Yuanming_ICC2015SDP}, which will require further investigation.

\item It is desirable to establish the optimality and perform probabilistic analysis for the iterative reweighted-$\ell_2$ algorithm in the context of inducing group sparsity in the multicast beamforming vectors. {\rev{However, with the complicated constraints set, this becomes much more challenging compared with the compressive sensing problems with simple constraints (e.g., affine constraints) \cite{Daubechies_2010iteratively}. It is also interesting to apply this algorithm to solve other mixed combinatorial optimization problems in wireless networks, e.g., wireless caching problems \cite{Jun_caching2014}.}} 

\item {\rev{It is interesting to apply the proposed smoothed $\ell_p$-minimization approach with the iterative reweighted-$\ell_2$ algorithm in the scenarios with CSI uncertainty. The only requirement is that the resulting non-convex constraints due to the CSI uncertainty can be convexified \cite{shen2012distributed, Yuanming_TSP14SCB}. It is also interesting but challenging to characterize the performance degradations due to CSI acquisition errors in multicast Cloud-RANs \cite{caire2010multiuser, Aubry_TIT}.}}  
\end{itemize}

\appendices
\section{Proof of Theorem {\ref{kkt}}}
\label{kktap}
We first need to prove that
\begin{eqnarray}
\label{sup1}
\limsup_{\epsilon\searrow 0}\Omega({\epsilon}) \subset\Omega({0}),
\end{eqnarray}
{\rev{where $\limsup_{\epsilon\searrow 0} \Omega(\epsilon)$ is defined as \cite[Page 152]{rockafellar1998variational}
\begin{eqnarray}
\limsup_{\epsilon\searrow 0}\Omega(\epsilon)&:=&\bigcup_{\epsilon^{[n]}\searrow 0} \limsup_{n\rightarrow \infty} \Omega(\epsilon^{[n]})\nonumber\\
&=& \{\bar{\bs{z}}|\exists \epsilon^{[n]}\searrow 0, \exists {\bs{z}}^{[n]}\rightarrow \bar{\bs{z}}\},
\end{eqnarray}
where ${\bs{z}}^{[n]}\in\Omega(\epsilon^{[n]})$. Therefore, for any $\bar{\bs{z}}\in\limsup_{\epsilon\searrow 0}\Omega({\epsilon})$, there exists ${\bs{z}}^{[n]}\in\Omega({\epsilon}^{[n]})$ such that ${\bs{z}}^{[n]}\rightarrow\bar{\bs{z}}$ and $\epsilon^{[n]}\searrow 0$}}. To prove (\ref{sup1}), we only need to prove that $\bar{\bs{z}}\in\Omega(0)$. Specifically, ${\bs{z}}^{[n]}\in\Omega({\epsilon}^{[n]})$ indicates that 
\begin{eqnarray}
\label{ktn}
 0\in \nabla_{\bs{z}}f_p({\bs{z}}^{[n]};
\epsilon^{[n]})+\mathcal{N}_{\mathcal{C}}({\bs{z}}^{[n]}).
\end{eqnarray}
As $f_{p}({\bs{z}}; \epsilon)$ is continuously differentiable in both $\bs{z}$ and $\epsilon$, we have
\begin{eqnarray}
\label{lim1}
\lim_{n\rightarrow\infty}\nabla_{\bs{z}}f_p({\bs{z}}^{[n]};
\epsilon^{[n]})&=&\lim_{{\bs{z}}^{[n]}\rightarrow\bar{\bs{z}}}\lim_{\epsilon^{[n]}\searrow0}\nabla_{\bs{z}}f_p({\bs{z}}^{[n]};
\epsilon^{[n]})\nonumber\\
&=&\nabla_{\bs{z}}f_p(\bar{\bs{z}};
0).
\end{eqnarray} 
Furthermore, based on results for the limits of normal vectors in \cite[Proposition 6.6]{rockafellar1998variational}, we have
\begin{eqnarray}
\label{lim21}
\limsup_{{\bs{z}}^{[n]}\rightarrow \bar{\bs{z}}}\mathcal{N}_{\mathcal{C}}({\bs{z}}^{[n]})=\mathcal{N}_{\mathcal{C}}(\bar{\bs{z}}).
\end{eqnarray}
{\rev{Specifically, if ${\bs{z}}^{[n]}\rightarrow\bar{\bs{z}}, {\bs{s}}^{[n]}\in\mathcal{N}_{\mathcal{C}}({\bs{z}}^{[n]})$ and ${\bs{s}}^{[n]}\rightarrow {\bs{s}}$, then ${\bs{s}}\in\mathcal{N}_{\mathcal{C}}(\bar{\bs{z}})$. That is, the set $\{({\bs{z}}, {\bs{s}})|{\bs{s}}\in\mathcal{N}_{\mathcal{C}}({\bs{z}})\}$ is closed relative to $\mathcal{C}\times \mathbb{R}^{m}$.}} Based on (\ref{lim1}) and (\ref{lim21}) and taking $n\rightarrow \infty$ in (\ref{ktn}), we thus prove (\ref{sup1}). Based on \cite[Theorem 4]{hong2011sequential}, we complete the proof for (\ref{kkt1}).

\section{Convergence of the Iterative Reweighted-$\ell_2$ Algorithm}
\label{converge}
1) We will show that any convergent subsequence $\{{\bs{z}}^{[n_k]}\}_{k=1}^{\infty}$ of $\{{\bs{z}}^{[n]}\}_{n=1}^{\infty}$ satisfies the definition of the KKT points of problem $\mathscr{P}_{\textrm{sm}}(\epsilon)$ (\ref{kktd}). Specifically, let ${\bs{z}}^{[n_k]}\rightarrow\bar{\bs{z}}$ be one such convergent subsequence with
\begin{eqnarray}
\label{limse}
\lim_{k\rightarrow\infty}{\bs{z}}^{[n_k+1]}=\lim_{k\rightarrow\infty}{\bs{z}}^{[n_k]}=\bar{\bs{z}}.
\end{eqnarray}
As 
\begin{eqnarray}
{\bs{z}}^{[n_k+1]}:=\arg\min_{{\bs{z}}\in\mathcal{C}}~Q({\bs{z}}; {\bs{\omega}}^{[n_k]}),
\end{eqnarray}
which is a convex optimization problem, the KKT condition holds at ${\bs{z}}^{[n_k+1]}$, i.e.,
\begin{eqnarray}
\label{kktse}
0\in \nabla_{\bs{z}}Q({\bs{z}}^{[n_k+1]}; {\bs{\omega}}^{[n_k]})+\mathcal{N}_{\mathcal{C}}({\bs{z}}^{[n_k+1]}).
\end{eqnarray}
Based on \cite[Proposition 6.6]{rockafellar1998variational} and (\ref{limse}),
we have
\begin{eqnarray}
\label{lim2}
\limsup_{{\bs{z}}^{[n_k+1]}\rightarrow \bar{\bs{z}}}\mathcal{N}_{\mathcal{C}}({\bs{z}}^{[n_k+1]})=\mathcal{N}_{\mathcal{C}}(\bar{\bs{z}}).
\end{eqnarray}
Furthermore, based on (\ref{limse}), we also have
\begin{eqnarray}
\lim_{k\rightarrow\infty}\nabla_{\bs{z}}Q({\bs{z}}^{[n_k+1]}; {\bs{\omega}}^{[n_k]})&=&\lim_{k\rightarrow\infty} 2\sum_{i=1}^{m}\omega^{[n_k]}z_i^{n_k+1}\nonumber\\
&=&\lim_{k\rightarrow \infty}\sum_{i=1}^{m}{{pz^{[n_k+1]}}\over{{{\left[\left(z_i^{[n_k]}\right)^2+\epsilon^2\right]^{1-{p\over{2}}}}}}}\nonumber\\
&=& \nabla_{\bs{z}}f_p(\bar{\bs{z}}; \epsilon).
\end{eqnarray}
Therefore, taking $k\rightarrow\infty$ in (\ref{kktse}), we have
\begin{eqnarray}
0\in \nabla_{\bs{z}}Q(\bar{\bs{z}}; \bar{\bs{\omega}})+\mathcal{N}_{\mathcal{C}}(\bar{\bs{z}}),
\end{eqnarray}
which indicates that $\bar{\bs{z}}$ is a KKT point of problem $\mathscr{P}_{\textrm{sm}}(\epsilon)$. We thus complete the proof.

2) As $f_{p}({\bs{z}};\epsilon)$ is continuous and $\mathcal{C}$ is compact, we have the fact that the limit of the sequence $f_{p}(\bs{z}^{[n]};\epsilon)$ is finite. Furthermore, we have $f_{p}({\bs{z}}^{[n+1]};\epsilon)\le f_{p}({\bs{z}}^{[n]};\epsilon)$ according to (\ref{ub1}). Based on the results in 1), we complete the proof. Note that a similar result was presented in \cite{Ba_2014convergenceTSP} by leveraging the results in the EM algorithm theory.

\bibliographystyle{ieeetr}

\end{document}